# On the localization transition in three dimensions: Monte-Carlo simulation of a non-linear $\sigma$-model

Thomas Dupré

*Institut für Theoretische Physik der Universität zu Köln*
*D-50937 Köln, Germany*

We present a combination of analytical and numerical calculations for the critical behavior of a supersymmetric non-linear $\sigma$-model. This model is expected to describe at least qualitatively the localization transition of a disordered one-electron system. As a result, we obtain a localization length exponent and a set of inverse participation numbers in three dimensions. We find a continuous phase transition with the features of one-parameter scaling and multifractality at the critical point.

## I. INTRODUCTION

The phenomenon of particle localization in a disordered medium has attracted a considerable interest starting with the famous Anderson paper [1]. The scaling theory [2,3] predicts a transition from extended to localized electronic states for any spatial dimension $d > 2$ and sufficiently large disorder. It assumes that this transition (Anderson transition, localization transition, metal-insulator transition (MIT)) can be described as a phase transition of second order with just one single relevant scaling field. Wegner [4] introduced the so-called $N$-orbital model which allowed for a field-theoretical reformulation of the problem [5–7] followed by a renormalization group analysis in $2+\epsilon$ dimensions [7–9]. It was recognized that a usual mean-field approach is not possible in this field theory since the conventional one-point Green function – which is closely related to the mean density of states (DOS) – does not exhibit a critical behavior. This seems to break the Goldstone theorem. [10] By extending and making rigorous this earlier work (which made use of the mathematically ill-defined replica trick) Efetov [11] mapped the problem of disordered one-electron systems on supersymmetric non-linear $\sigma$-models (in the following called supermatrix models) with certain super coset spaces $\mathbf{G}/\mathbf{K}$ according to the particular behavior of the disordered one-electron system under time reversal and/or spin rotation.

The investigation of the aforementioned supermatrix models on the Bethe lattice (BL) [12–14] revealed an unexpected behavior at the transition point. Namely, a *jump* in the inverse participation ratio (IPR) on the localized side and an *exponential* decrease of the diffusion constant on the metallic side of the transition was found. These results gave rise to the hypothesis [14] that the scaling theory (which predicts a power-law behavior of various quantities) might be not applicable for the localization transition. In order to investigate the critical behavior of this transition on a hypercubic lattice, a Lagrangian was constructed in Ref. [15] so that its saddle-point reproduced the so-called effective-medium approximation (EMA) of Ref. [16] which leads to the same non-power-like critical behavior as in the case of the BL. It was proposed [17] that this exotic behavior is characteristic for the Anderson transition on a $d$-dimensional hypercubic lattice for a (sufficiently high) spatial dimension $d > 2$.

However, Mirlin and Fyodorov [18] argued that contrary to the above hypothesis the non-power-like critical behavior is an artifact of the special lattice structure of the BL and of the EMA which imitates this structure. One can assign an effective dimension $d = \infty$ to this lattice structure, which plays the role of the upper critical dimension. (A transition behavior with critical exponents 0 and $\infty$ is then formally identified with a jump and an exponential behavior, resp.) In this picture the above critical behavior on the BL or within the EMA can be understood within the one-parameter scaling assumption; for finite dimensions a power-like scaling behavior is expected.

Moreover, it was shown in a quantitative way [19] that the EMA is an uncontrolled approximation due to the neglect of loop graphs, since in the context of a strong-coupling expansion the highly branched graphs with a high number of loops yield the dominant contribution whereas zero-loop graphs can be neglected. It is therefore perhaps fair to say that instead of trying to calculate corrections to the BL results it seems more promising to investigate the localization transition problem (within the supersymmetry formalism) directly in $d = 3$ dimensions, at least numerically. This has, however, not yet been done.

The development of a supersymmetric toy model, the Hyperbolic Superplane (HSP) [20], allowed for some tech-nical simplifications in the investigations of non-linear $\sigma$-models. It was shown in [20] that this model is capable of describing localization in a quasi one-dimensional geometry. On the BL, an analysis of a non-linear $\sigma$-model taking the HSP as target space [19] reproduced the previously reported unconventional critical behavior. This experience indicates that the HSP might yield a qualitatively correct picture of the localization transition in $d > 2$ dimensions, too.

Technical simplifications compared with the aforementioned supermatrix spaces originate from the fact that the HSP has only *one* radial coordinate whereas the supermatrix models show at least two radial degrees of freedom. It may thus serve as a useful toy model which allows one to study many interesting features related to the MIT with relative ease, thereby gaining some experience and insight into the main difficulties before one turns to the more interesting supermatrix spaces. (The HSP has been successfully applied to several other problems as well, e.g. Migdal-Kadanoff renormalization [21] or quantum chaos in conjunction with a superanalog of the Selberg trace formula [22].)

Thus, the goal of the present paper is to help clarify the above controversy by a mainly numerical investigation (Monte-Carlo simulation) of a non-linear $\sigma$-model in $d = 3$ dimensions taking the HSP as coset space $\mathbf{G}/\mathbf{K}$. To anticipate the main result, we find a continuous phase transition, one-parameter scaling (1PS) and multifractality. In other words, the data yield no evidence for a jump of the IPR. The localization length exponent of our model is obtained as $\nu = 1.15 \pm 0.15$.

This work is organized as follows: After a short introduction of the HSP in section II we state in section III the model which is to be investigated. In section IV we describe the main ideas which enable us eventually to treat the problem using a Monte-Carlo (MC) algorithm. The results and fundamental limitations of this algorithm are outlined as well. We conclude with some remarks concerning the applicability of the MC algorithm for related problems.

## II. THE HYPERBOLIC SUPERPLANE (HSP)

### A. Geometry of the HSP

The HSP is a supersymmetric homogeneous space $\mathbf{G}/\mathbf{K}$ of rank one and real dimension (2,2) and can be viewed as a real, supersymmetric extension of the (upper) hyperboloid $H_2 = SO(2,1)/SO(2)$. It shares several important aspects with Efetov's supermatrix spaces [11], more precisely, with the model I for the case of orthogonal symmetry. These are (besides the reality constraint)

- its structure as coset space $\mathbf{G}/\mathbf{K}$
- perfect grading, i.e. equal numbers of bosonic (commuting, even) and fermionic (anticommuting, odd) degrees of freedom
- non-compactness
- a positive (super-)Riemannian curvature

A detailed introduction of the HSP – in particular, a precise definition as a homogeneous space $\mathbf{G}/\mathbf{K}$– was given in Ref. [20] (see also Ref. [21]); the complex counterpart of the real HSP was introduced in [22]. These definitions will not be repeated here. Instead, for our purposes it will be sufficient to give a brief idea of the HSP and to put forward a useful parameterization.

An element $\psi$ of the real HSP can be viewed as a five-component supervector, $\psi^T = (\psi_0, \psi_1, \psi_2, \psi_3, \psi_4)$ where $\psi_0, \psi_1, \psi_2$ are even and $\psi_3, \psi_4$ are odd. We will use an involution (adjoint, complex conjugation) of the second kind which means that for odd elements $\xi_i, \xi_j$ we define

$$\overline{(\overline{\xi_i})} = -\xi_i \quad \text{and} \quad \overline{(\xi_i \xi_j)} = \bar{\xi}_i \bar{\xi}_j \; .$$

Using this involution we can express the reality condition for the HSP as

$$\psi_0, \psi_1, \psi_2 \in \mathbf{R} \; , \quad \overline{\psi_3} = \psi_4 \; .$$

The supervector $\psi$ obeys a (non-linear) constraint

$$\psi^\dagger \theta \psi = 1$$

where $\psi^\dagger = \overline{\psi}^T$ and the metric $\theta$ is given by

$$\theta = \text{diag}(1, -\mathbf{1}_4) \; . \tag{1}$$

Moreover, we require for the body [1] $m(\psi_0)$ of $\psi_0$ the condition $m(\psi_0) \geq 1$. We now introduce polar coordinates,

$$\psi = \begin{pmatrix} \psi_0 \\ \vdots \\ \psi_4 \end{pmatrix} = \begin{pmatrix} \cosh r \\ \sinh r \; e \end{pmatrix} \; , \; e = \begin{pmatrix} \cos\phi \; (1 - \bar{\eta}\eta) \\ \sin\phi \; (1 - \bar{\eta}\eta) \\ \eta \\ \bar{\eta} \end{pmatrix} \tag{2}$$

with $r, \phi$ even, $0 \leq r$, $0 \leq \phi < 2\pi$ and odd elements $\eta, \bar{\eta}$. Let us denote the "radial" part of a supervector $\psi$ by $\psi^r$,

$$\psi^{rT} = (\cosh r, \sinh r, 0, 0, 0) \; , \tag{3}$$

and the origin of the HSP by $o$,

$$o^T = (1, 0, 0, 0, 0) \; . \tag{4}$$

(The above coordinates are motivated on a group-theoretical level in Ref. [20].)

---

[1] The body $m(x)$ of an even element $x$ denotes the ordinary part of $x$, which results after subtraction of all nilpotent terms of $x$.

## III. THE MODEL

### A. Definition and motivation of the model

In this section we introduce and motivate the model which is to be investigated; we also define some correlation functions of interest and specify the order parameter function. Finally we show how physical quantities can be extracted from the model.

The model Efetov used [14] is a system of impure metal granules (with a large number of states) which are in contact with each other. The (macroscopic) conductivity is then essentially governed by the probability of the electron jumping from one granule to a neighboring one. This "granular" model is closely related to Wegner's $N$-orbital model [4]. There is some evidence [23] that the Anderson model and the $N$-orbital model (or the granular model) exhibit the same critical behavior. The above model of disordered metal granules can be described by a supersymmetric non-linear $\sigma$-model on a lattice; its free energy (for identical granules) is written in the form [14]

$$F[Q] = -J \sum_{i,j} \text{str} Q_i Q_j - \beta \sum_i \text{str} \Lambda Q_i . \quad (5)$$

Here, the first sum runs over nearest-neighbor granules, $J$ is related to the coupling between the granules and the parameter $\beta$ is given by $\beta = \frac{i}{4}(\omega + i\epsilon)\pi\nu V$, where $V$ denotes the volume of a granule, $\nu$ the average DOS, $\epsilon \to 0^+$ and $\omega$ is a frequency. The supermatrices $Q$ and $\Lambda$ have dimension $8 \times 8$. Their precise forms can be found in [11,14].

The correlation function of the DOS, $K_1$, and the density-density correlator, $K_2$, are given by

$$K_1(x, y; E) = \overline{G^+(x,x;E) G^-(y,y;E)} , \quad (6)$$
$$K_2(x, y; E; \omega) = \overline{G^+(x,y;E) G^-(y,x;E-\omega)} , \quad (7)$$

where the overbar stands for the disorder average and $G^+$ and $G^-$ are the retarded and advanced Green functions. These correlators can be calculated via

$$K_1(x, y; E) \propto \int \prod_i dQ_i \ Q^{11}_{11}(x) Q^{22}_{11}(y) \exp -F[Q], \quad (8)$$

$$K_2(x, y; E; \omega) \propto \int \prod_i dQ_i \ Q^{12}_{13}(x) Q^{21}_{31}(y) \exp -F[Q]. \quad (9)$$

Here, upper indices denote $4 \times 4$ sub-matrices of the $Q$-matrices and lower indices the corresponding entries within these blocks.

Now let us consider the following statistical toy model: At each site $x$ of a given lattice attach a field $\psi(x)$ where $\psi$ is an element of the HSP. We define the action (resp. the free energy) of the model as follows:

$$S[\psi] = \gamma \sum_{x \smile y} (\psi^\dagger(x) \theta \psi(y) - 1) + \beta \sum_x (\psi_0(x) - 1)$$
$$\equiv S_0[\psi] + S_1[\psi] , \quad (10)$$
$$\beta = -i(\omega + i\epsilon)\nu V , \quad \epsilon \to 0^+ ,$$

where $\theta$ was defined in eq. (1) and the notation $x \smile y$ means here and henceforth that $x$ and $y$ are nearest neighbors. The "interaction term" $S_0[\psi]$ contains a sum over all nearest neighbor pairs with an inverse coupling constant $\gamma > 0$ which is a measure for the disorder of the system. Of central importance are the symmetry properties of $S[\psi]$: The term $S_0$ is invariant under a "global" transformation $\psi(x) \mapsto g \cdot \psi(x)$, $g \in \mathbf{G}$ . The second term, $S_1$, breaks this symmetry, but it leaves a residual symmetry unbroken. Namely, $S_1$ is invariant under the action of the stability group $\mathbf{K} \subset \mathbf{G}$, i.e. under global transformations of the form $\psi(x) \mapsto k \cdot \psi(x)$, $k \in \mathbf{K}$. (The action of elements of the stability group $\mathbf{K}$ leaves the origin (4) invariant.) Expectation values of functions $A(\psi)$ are defined in the usual way,

$$< A(\psi) > \ = Z^{-1} \int [D\psi] A(\psi) \exp -S[\psi] ,$$
$$[D\psi] = \prod_{i=1}^N D\psi(x_i) .$$

Here, $N$ is the number of lattice sites, and for the partition function $Z$ we have

$$Z = \int [D\psi] \exp -S[\psi] = \exp -S[\psi]|_{\psi(x)=o} = 1$$

because of the $\mathbf{K}$-invariance which remains unbroken.

In order to motivate the present model (10) we note that a detailed analysis of the supermatrix models of the form (5) has revealed [13,14] that the *critical* properties of these models depend crucially on their symmetry structure with respect to the groups $\mathbf{G}$ and $\mathbf{K}$. The toy model (10) on the HSP imitates exactly this symmetry structure and is therefore expected to yield a qualitatively correct description of the critical behavior of the localization transition.

The use of the HSP which has only one (non-compact) radial coordinate is further motivated by the observation that the critical behavior of the order parameter function (see below) is governed by just this non-compact variable; the other (compact) radial coordinate(s) of the supermatrix models are of no interest in this case. This allows for a description of the MIT using the simplified toy model. (Of course, one cannot hope to use this toy model successfully in situations where the compact radial sector of the supermatrix spaces is of equal importance as the non-compact one.)

## B. Correlation functions and physical quantities

Notwithstanding the formal analogy between the model (10) and spin models, some usual methods which are used in phase transition theory are not applicable here. These include in particular mean-field methods, which involve the magnetization, which is non-vanishing only in the ferromagnetic phase and identically zero in the paramagnetic phase. The reason is the equality $<\psi>= o$ which follows from the unbroken **K**-invariance so that the mean DOS is clearly not critical. (This is in accordance with the situation in the supermatrix models where the expectation value of $Q$ equals $\Lambda$ and is therefore a constant.)

The only non-trivial 2-point function of our model is

$$K(x,y;\gamma,\beta) = K(x-y;\gamma,\beta) = (-)^{|\mu|} < \overline{\psi_\mu}(x)\psi_\mu(y) > \quad (11)$$

with

$$\mu \in 1,\ldots,4 \quad , \quad |\mu| = \begin{cases} 0 & , \; \psi_\mu \text{ even} \\ 1 & , \; \psi_\mu \text{ odd} \end{cases} .$$

(Note that $<\overline{\psi_0}(x)\psi_0(y)>$ is not critical. This is a consequence of the fact that the considered model has less degrees of freedom than Efetov's supermatrix models.)

The value of the coupling constant $\gamma$ decides whether the system is in the phase of localized or extended states (if both phases exist). Let $\gamma_c$ be the value of $\gamma$ at which a possible phase transition occurs. Then we have the correspondence

$$\text{localized states} \leftrightarrow \gamma < \gamma_c$$
$$\text{extended states} \leftrightarrow \gamma > \gamma_c$$

Both phases differ from each other with respect to the symmetry breaking behavior [13]. In the disordered phase (insulating phase, phase of localized states) the broken **G**-invariance is restored in the limit $\beta \to 0$. The conducting phase (metallic phase, phase of extended states), however, exhibits a scenario of spontaneous symmetry breaking: the **G**-invariance remains broken even after the symmetry breaking term $S_1$ vanishes in the limit $\beta \to 0$. As was shown in Ref. [13] in the context of the supermatrix models this different symmetry behavior together with the non-compactness of **G/K** implies a different behavior of the correlation functions. One finds

$$\lim_{\beta \to 0} \beta K(x-y;\gamma,\beta) = \begin{cases} const & , \; \text{localized states} \\ 0 & , \; \text{extended states} \end{cases} \quad (12)$$

The two phases of the system can be distinguished by an order parameter, more precisely by an order parameter function. [13] Let $F(\psi(x))$ be that function which is obtained if one integrates the statistical weight $\exp -S[\psi]$ over all sites of an (infinitely extended) lattice except the site $x$,

$$F(\psi(x)) = \int \prod_{y \neq x} D\psi(y) \exp -S[\psi] . \quad (13)$$

For symmetry reasons $F$ depends only on the radial part $\psi^r$ (3) of $\psi$ and can be used to discriminate the two phases,

$$\lim_{\beta \to 0} F(\psi^r) \propto \begin{cases} 1 \text{ (the constant function)} & , \; \gamma < \gamma_c \\ \text{a for } r \to \infty \text{ decaying function} & , \; \gamma > \gamma_c \end{cases}$$

A physical meaning of the above function $F$ was given in [18,23] where it was shown that $F$ is closely related to the distribution function of the local densities of states (LDOS).

For future reference let us define

$$L(\psi,\psi') = \exp(-\gamma(\psi^\dagger \theta \psi' - 1)) , \quad (14)$$
$$D_\beta(\psi) = \exp(-\beta(\psi_0 - 1)) . \quad (15)$$

These functions show the following important symmetries,

$$L(\psi,\psi') = L(g\psi, g\psi') \quad \forall g \in \mathbf{G} ,$$
$$D_\beta(\psi) = D_\beta(k\psi) \quad \forall k \in \mathbf{K} ,$$

which means that $D_\beta$ is a **K**-radial function,

$$D_\beta(\psi) = D_\beta(\psi^r) \equiv D_\beta(r) .$$

In addition to the correlator (11), we will also frequently be interested in "diagonal" correlators of the type

$$K^q(\gamma,\beta) = \mathcal{N}_{l,m}^{-1} \left\langle \psi_1^{\;l}(x)\psi_2^{\;m}(x) \right\rangle , \quad (16)$$
$$\mathcal{N}_{l,m} = (l+m)\frac{1}{\pi}\frac{\Gamma(\frac{l+1}{2})\Gamma(\frac{m+1}{2})}{\Gamma(\frac{l+m+2}{2})}(l+m-2)! ,$$

where $l, m$ are integers larger than or equal to zero, $q = l+m \geq 2$, and $x$ is an arbitrary lattice site. The normalization factor is chosen so that the dependence of $K^q$ on $l$ and $m$ enters only via the sum $q = l+m$ and that for $\gamma \to 0$ $K^q$ approaches unity. [For $q=2$ and $l=2, m=0$ (or vice versa) this correlator reduces – since $\psi_1$ and $\psi_2$ are real – to the correlation function $K(0;\gamma,\beta)$, eq. (11).]

Let us now demonstrate how we can (at least formally) extract physical quantities from the above defined correlation functions. Motivated by the unusual properties of the MIT on the BL we will focus our interest mainly onto the following quantities: the IPR (and its higher moments) and the diffusion coefficient $D$.

Let us start from a tight-binding model with lattice sites $x, y, \ldots$, normalized eigenstates $\psi_k$ and an average level density $\overline{\rho}(E)$ per site at energy $E$. The inverse participation ratios (IPR's) $P_q(E), q \in \mathbb{R}$,

$$P_q(E) = \overline{\int d^d r |\psi(r)|^{2q}} = \frac{1}{\overline{\rho}(E)} \sum_k \overline{|\psi_k(x)|^{2q} \, \delta(E - E_k)},$$

are a very sensitive measure for the degree of localization in the system. [24] To which extent the states are localized, can be inferred from the dependence of $P_q$ on the length $L$ of the system (we set $N = L^d$, where $d$ is the spatial dimension). For extended states (deep in the metallic region), $P_q$ scales like

$$P_q \propto N^{-(q-1)} \propto L^{-(q-1)d} \quad (17)$$

whereas for strong disorder (which means that the localization length $\xi$ is much smaller than $L$) $P_q$ is rather independent of $L$. In the limit of extremely localized states all $P_q$ tend towards unity. Thus, the IPR's can be used to discriminate localized from extended states.

Around the critical point $\gamma_c$ (i.e. in the range where $\xi$ exceeds by far any microscopic length scale $l_{mic}$) one makes the following ansatz relying on the assumption of one-parameter scaling (1PS),[2]

$$P_q(\gamma; L) \propto L^{-\tau(q)} f_q(\frac{L}{\xi}), \quad \tau(q) \equiv (q-1)d^*(q) \quad (18)$$

with some scaling functions $f_q$. The "generalized dimension(s)" $d^*(q)$ differs from the spatial dimension $d$ and is a function of $q$. For $L \to \infty$ $P_q$ must become independent of $L$, therefore the scaling functions $f_q$ must (for $L \gg \xi$) satisfy $f_q(L/\xi) \propto (L/\xi)^{\tau(q)}$. Thus, $P_q(\gamma \simeq \gamma_c; L \to \infty) \propto \xi^{-\tau(q)}$. Using the relation $\xi \propto (\gamma_c - \gamma)^{-\nu}$ (for $\gamma \lesssim \gamma_c$) one concludes therefore that $P_q$ scales like

$$P_q(\gamma \lesssim \gamma_c; L \to \infty) \propto (\gamma_c - \gamma)^{\pi(q)}, \quad \pi(q) = \nu\tau(q), \quad (19)$$

and the multifractality of the system is reflected in a non-trivial dependence of $\pi(q)$ on $q$. One can show in a rather general context (see for example [25] and references herein) that $\tau(q)$ is a monotonically increasing function with negative curvature.

The IPR's can be rewritten in terms of Green functions, [24,26]

$$P_2(E) = \lim_{\eta \to 0} \frac{\eta}{\pi \overline{\rho}(E)} \overline{G^+(x,x;E)G^-(x,x;E)},$$

$$P_q(E) = \lim_{\eta \to 0} \frac{i^{l-m}(2\eta)^{q-1}}{2\pi\overline{\rho}} \frac{(l-1)!(m-1)!}{(l+m-2)!} \times$$
$$\times \overline{(G^+(x,x;E))^l \, (G^-(x,x;E))^m}$$

where $q = l + m$ and $l, m \in \mathbf{N}$.

Using in a second step the techniques of supersymmetry, the above averaged products of Green functions can be represented in terms of certain correlators, i.e. expectation values of the bosonic blocks of the $Q$-matrices with respect to the generating functional $F(Q)$, see for instance equations (6)-(9) and Ref. [26]. Because our model (10) is only a toy model there exists, however, no such immediate physical interpretation of the correlators $K$ (11) and $K^q(\gamma; \beta)$ (16). Nevertheless, $K$ corresponds formally to the correlator (9), which in turn is related to the density-density correlation function $K_2(x, y; E; \omega)$, eq. (7). Further, the identity [13]

$$\lim_{\eta \to 0} \eta K_1(x, y; E) = \lim_{\eta \to 0} \eta K_2(x, y; E; 0)$$

relates the correlation functions (6) and (7) in the localized regime. Therefore $K$ plays *also* the role of $K_1$, and we obtain

$$P_2(\gamma) = \lim_{\beta \to 0} \beta K(0; \gamma, \beta) . \quad (20)$$

Similarly, concerning the whole set of IPR's ($q \geq 2$), we can formally identify

$$P_q(\gamma) = \lim_{\beta \to 0} \beta^{q-1} K^q(\gamma, \beta) \quad (21)$$

which relates (at least, formally) the IPR's $P_q$ to expectation values of the model (10). Let us note, however, that due to the reduced number of variables in our toy model (compared with the supermatrix spaces) the correlator $K^q$ is non-trivial only for *even* $l$ and $m$. Thus, we can only hope to calculate IPR's with an even $q$ within our model.

The diffusion coefficient $D(\gamma)$ can be extracted by comparison with the long-range behavior of the correlator (7) – or, equivalently, (11) – on the conducting side of the transition. More precisely, its Fourier transform reads

$$K(k, \omega; \gamma) \propto \frac{1}{D(\gamma)k^2 - i\omega}, \quad k \to 0, \omega \to 0 \quad (22)$$

from which $D(\gamma)$ can be inferred. Consequently, we can calculate $P_q(\gamma)$ and $D(\gamma)$ by consideration of the *single* function $K$.

### C. A simple illustration

The simplest case, which can be considered, is a lattice with just $N = 2$ lattice sites, $x$ and $y$, with attached fields $\psi$ and $\psi'$, resp. Let us evaluate the correlator (21) and the IPR's (21) for this particular case. We have

$$P_q(\gamma) = \mathcal{N}_{l,m}^{-1} \lim_{\beta \to 0} \beta^{q-1} \int D\psi \, \overline{\psi_1}^l \psi_2^m F(\psi^r),$$

$$F(\psi^r) = D_\beta(\psi^r) \int D\psi' \, L(\psi^r, \psi') D_\beta(\psi').$$

We can use polar coordinates (2) and easily integrate out the variables $(\phi, \eta, \overline{\eta})$ at the site $x$ which yields (for $l, m$ even and greater than or equal to zero) a factor $\frac{\mathcal{N}_{l,m}}{(l+m-2)!}$. In order to integrate out the remaining variables note that because of the infinitesimal $\beta^{q-1}$ only the asymptotic domains $r, r' \to \infty$ contribute ($\cosh r, \cosh r' \sim \beta^{-1}$), hence we substitute

$$\omega = \beta \cosh r, \omega' = \beta \cosh r', \varphi' = \phi'/\beta, \chi' = \eta'/\beta . \quad (23)$$

---

[2]Of course, this 1PS ansatz has to be checked explicitly by a numerical investigation.

$P_q(\gamma)$ becomes then, using standard integral identities [27],

$$P_q(\gamma) = \lim_{\beta \to 0} \frac{\beta^{q-1}}{(q-2)!} \int_0^\infty \frac{d\omega}{\beta} \left(\frac{\omega}{\beta}\right)^{q-2} e^{-\omega} \int_0^\infty \frac{d\omega'}{\beta}$$
$$\left(\frac{\beta}{\omega'}\right)^2 e^{-\omega'} \int \beta d\varphi' \frac{\partial_{\bar{\chi}'} \partial_{\chi'}}{\beta^2} L_0(\omega, \omega', \varphi')$$
$$(1 - \gamma \omega \omega' \bar{\chi}' \chi')$$
$$= \int_0^\infty \frac{d\omega \; \omega^{q-2}}{(q-2)!} \exp\left[\gamma - \omega - 2\sqrt{\frac{\gamma}{2}(\omega + \frac{\gamma}{2})}\right] \quad (24)$$

with

$$L_0(\omega, \omega', \varphi') = \exp -\gamma \left[\frac{1}{2}(\frac{\omega}{\omega'} + \frac{\omega'}{\omega}) - 1 + \frac{1}{2}\omega \omega' \varphi'^2\right]. \quad (25)$$

Note that all powers of $\beta$ have cancelled. After a straightforward calculation one obtains for the last remaining integral

$$P_q(\gamma) = \frac{(-)^{q-2} \gamma^{q-2}}{(q-2)!} (\partial_\gamma)^{q-2} \times$$
$$\left[1 - \delta \sqrt{\frac{\pi}{2\gamma}} e^{(\gamma+\delta)^2/2\gamma} \left(1 - \phi(\frac{\gamma+\delta}{\sqrt{2\gamma}})\right)\right]_{\delta=\gamma}$$

where $\phi$ is the error function. In particular, for $q = 2$ the result is

$$P_2(\gamma) = 1 - \sqrt{\frac{\pi \gamma}{2}} e^{2\gamma} \left(1 - \phi(\sqrt{2\gamma})\right) .$$

The limiting behavior of $P_q(\gamma)$ for $\gamma \to 0$ and $\gamma \to \infty$ can easily be read off,

$$P_q(\gamma) \to \begin{cases} 1 & , \gamma \to 0 \\ 2^{-(q-1)} & , \gamma \to \infty \end{cases}$$

as required in eq. (17).

Let us now turn our attention to the three-dimensional case where integrals of the above type can, of course, be solved only numerically.

## IV. MONTE-CARLO SIMULATION

In this section we introduce a MC algorithm which allows us to calculate some correlation functions of the model (10) numerically. For simplicity, we will restrict ourselves to *diagonal* correlation functions; specifically we will calculate the set of IPR's $P_q(\gamma)$ (21). In principle the algorithm is capable of calculating non-diagonal correlation functions as well.

In order to apply MC techniques it is absolutely necessary, however, to find a method which treats the fermionic variables appropriately. It is well-known that attempts to study systems which include dynamical fermions by means of MC simulations face severe problems. This is mainly due to the so-called "minus sign" problem which arises from the presence of Grassmann variables that make a definition of a positive-definite probability measure problematic [28]. However, it turns out that an MC simulation of the model (10) is possible in spite of the presence of fermionic variables, because of the self-terminating property of the Grassmann polynomials.

The underlying idea of the algorithm has been suggested some years ago [13] but has never been pursued. Thus, no numerical simulations of the kind below have yet been performed.

### A. Derivation of the Algorithm

To begin with, let $x$ be an arbitrary lattice site with attached field $\psi$. We write the IPR $P_q(\gamma)$ (21) in the following form

$$P_q(\gamma) = \mathcal{N}_{l,m}^{-1} \lim_{\beta \to 0} \beta^{q-1} \int D\psi \; \overline{\psi_1}^l \psi_2^m F(\psi^r) ,$$
$$F(\psi^r) = D_\beta(\psi^r) \int \prod_{y \neq x} D\psi(y) D_\beta(\psi^r(y)) \quad (26)$$
$$\prod_{x' \frown x} L(\psi^r, \psi(x')) \prod_{x_i, x_j \neq x} L(\psi(x_i), \psi(x_j))$$

where $q = l + m$. ($P_q(\gamma)$ does not depend on the specific choice of $x$, of course.) Similar as in section III C we integrate out the angular and fermionic variables at the site $x$. The substitution $\omega = \beta \cosh r$ leads afterwards to [3]

$$P_q(\gamma) = \lim_{\beta \to 0} \int d\omega(x) e^{-\omega(x)} \frac{\omega(x)^{q-2}}{(q-2)!} \int \prod_{y \neq x} D\psi(y) e^{-\omega(y)}$$
$$\prod_{x' \frown x} L(\psi^r(x), \psi(x')) \prod_{x_i, x_j \neq x} L(\psi(x_i), \psi(x_j))$$
$$= Z^{-1} \left\langle \left\langle \frac{\omega(x)^{q-2}}{(q-2)!} \prod_{y \neq x} e^{-\omega(y)} \right\rangle \right\rangle$$

---

[3] In the limit $\beta \to 0$ no boundary terms arise when one changes to polar coordinates. – The $\omega$-integrations run from 0 to $\infty$.

with

$$\langle\langle\ldots\rangle\rangle \equiv \int d\omega(x) e^{-\omega(x)} I(\psi^r(x)) (\ldots),$$

$$I(\psi^r(x)) = \int \prod_{y \neq x} D\psi(y) \prod_{x' \smile x} L(\psi^r(x), \psi(x'))$$
$$\prod_{x_i, x_j \neq x} L(\psi(x_i), \psi(x_j)).$$

Note that the normalization of this average is

$$Z = \langle\langle 1 \rangle\rangle = \int d\omega(x) e^{-\omega(x)} I(\psi^r(x)) = 1$$

because $I(\psi^r(x)) = 1$ due to the absence of symmetry breaking terms.

Thus, the correlation functions can be expressed as "thermodynamic" expectation values, similar to the concept of statistical mechanics. Thus, the following scheme becomes apparent:

- Generate configurations according to a certain distribution function which will be specified below.

- Average the symmetry breaking terms

$$\frac{\omega(x)^{q-2}}{(q-2)!} \prod_{y \neq x} e^{-\omega(y)}$$

    over these configurations.

Next comes the problem of how to deal with the Grassmann variables. Performing at each lattice site the substitution (compare (23))

$$(r, \phi, \bar\eta, \eta) \mapsto (\omega = \beta \cosh r, \varphi = \phi/\beta, \bar\chi = \bar\eta/\beta, \chi = \eta/\beta)$$

we integrate out all remaining Grassmann variables $(\bar\chi, \chi)$: First note that as a direct generalization of eq. (25) we find

$$L(\psi_i, \psi_j) = L_0(\omega_i, \omega_j, \varphi_i - \varphi_j)(1 - \gamma \omega_i \omega_j \overline{\Delta\chi_{ij}} \Delta\chi_{ij}) \tag{27}$$

where $\Delta\chi_{ij} \equiv \chi_i - \chi_j$. Now let us call a bond between $\psi_i$ and $\psi_j$ $\left\{\begin{array}{c}\text{non-occupied}\\\text{occupied}\end{array}\right\}$, if the "interaction term" according to eq. (27) is given by $\left\{\begin{array}{c}L_0\\-\gamma\omega_i\omega_j\overline{\Delta\chi_{ij}}\Delta\chi_{ij}\ L_0\end{array}\right\}$. Then, after integrating out the Grassmann variables, one can express $I(\psi^r(x))$ as a sum over all possible graphs (consisting of occupied and unoccupied bonds) on the underlying lattice. The following simple graphical rules for the evaluation of $I(\psi^r(x))$ hold:

1. *Each graph which contains a loop of occupied bonds, vanishes.* The reason for this is the identity

$$\Delta\chi_{12}\Delta\chi_{23}\ldots\Delta\chi_{n1} = 0 \ .$$

Thus, only tree graphs contribute to $I(\psi^r(x))$ .

2. *Each graph that is not linked to the site $x$ (via a cluster of occupied bonds) vanishes.* The reason for this is that for a non-vanishing contribution the number of Grassmann variables (provided by the occupied bonds) must match exactly the number of fermionic integrations (provided by the vertices *except* by the site $x$).

Thus, only tree graphs that emerge out of the site $x$, can contribute. Let us denote this set of contributing tree configurations by $\{G\}$.

After we have integrated out the Grassmann variables, the partition sum can be rewritten as

$$Z = \sum_{\{G\}} \int d\omega(x) e^{-\omega(x)} \int \prod_{y \neq x} \frac{d\omega(y)}{\omega^2(y)} \frac{d\varphi(y)}{2\pi}$$
$$\prod_{x_i \smile x_j} L_0(\omega_i, \omega_j, \varphi_i - \varphi_j) \prod_{x_k \smile x_l}^{\in\{G\}} \gamma\omega_k\omega_l.$$

In order to circumvent the - from a numerical point of view - inconvenient condition $\omega \geq 0$ together with the singularity of the measure at $\omega = 0$ the following substitution appears useful,

$$\lambda(x) = \exp -\omega(x) \ , \ t(y) = \ln \omega(y) \ (\text{for } y \neq x).$$

Thus, we have

$$P_q(\gamma) = \sum_{\{G\}} \int D[\phi] A_q(\phi) V(\phi, G),$$

$$\int D[\phi] = \int_0^1 d(e^{-\omega(x)}) \prod_{y \neq x} \int_{-\infty}^{\infty} d(\ln \omega(y)) \int_{-\infty}^{\infty} \frac{d\varphi(y)}{2\pi},$$

$$A_q(\phi) = \frac{\omega(x)^{q-2}}{(q-2)!} \prod_{y \neq x} \exp -\omega(y)$$

$$V(\phi, G) = \gamma^{N-1} \omega(x)^{z(x)} \prod_{y \neq x} \omega(y)^{z(y)-1}$$
$$\prod_{x_i \smile x_j} L_0(\omega_i, \omega_j, \varphi_i - \varphi_j),$$

where $z(x_k)$ is the local coordination number of the site $k$, i.e. the number of occupied bonds at $k$.

In principle one could now calculate $P_q(\gamma)$ via an MC algorithm on the configuration space $\{(\omega_i, \varphi_i), \{G\}\}$. However, such simulations have revealed that the relaxation into equilibrium is – even upon small lattices – very slow, and far from being satisfactory. It is therefore expedient to reduce the number of degrees of freedom of the model further, by integrating out the angular variables $\varphi$. This integration is just an $(N-1)$-dimensional Gaussian integral with the additional constraint $\varphi(x) = 0$ and yields

$$\prod_{k \neq x} \int_{-\infty}^{\infty} \frac{d\varphi_k}{2\pi} \exp\left(-\frac{\gamma}{2} \sum_{i \smile j} \omega_i \omega_j (\varphi_i - \varphi_j)^2\right)\Bigg|_{\varphi_x = 0} =$$

$$\left(\prod_{k\neq x}\omega_k^{-1}\right)(2\pi\gamma)^{-\frac{N-1}{2}}(\det A_x)^{-\frac{1}{2}}.$$

Here, the matrix $A_x$ results from the real, symmetric, positive definite matrix $A$ with

$$A_{ij} = \begin{cases} a_i^{-1}, & i=j \\ -1, & i \smile j \\ 0, & \text{otherwise} \end{cases}, \quad a_i = \frac{\omega_i}{2\sum_{j\smile i}\omega_j} \quad (28)$$

by erasing the line and column corresponding to the lattice site $x$. (Note that $\det A_x > 0$.) Therefore we can reduce the configuration space to $([\Omega]=\{\omega_1,\ldots,\omega_N\},\{G\})$, and obtain finally

$$P_q(\gamma) = \sum_{\{G\}} \int D[\Omega] A_q(\Omega) W(\Omega,G) \quad (29)$$

$$\int D[\Omega] = \int_0^1 d(e^{-\omega(x)}) \prod_{y\neq x} \int_{-\infty}^{\infty} d(\ln\omega(y)) \quad (30)$$

$$A_q(\Omega) = \frac{\omega(x)^{q-2}}{(q-2)!} \prod_{y\neq x} \exp -\omega(y) \quad (31)$$

$$W(\Omega,G) = \left(\frac{\gamma}{2\pi}\right)^{\frac{N-1}{2}} (\det A_x)^{-\frac{1}{2}} \omega(x)^{z(x)}$$
$$\prod_{y\neq x} \omega(y)^{z(y)-2} \prod_{i\smile j} L_\gamma(\omega_i,\omega_j) \quad (32)$$

$$L_\gamma(\omega_i,\omega_j) = \exp -\frac{\gamma}{2}\frac{(\omega_i-\omega_j)^2}{\omega_i\,\omega_j} \quad (33)$$

Notice that $\sum_{\{G\}} \int D[\Omega] W(\Omega,G) = 1$ and that the appearance of the exponents $z(y) - 2$ is "natural" because the average (local) coordination number $\bar{z}$ of a tree configuration is given by

$$\bar{z} = 2\frac{N-1}{N} \stackrel{N\to\infty}{\longrightarrow} 2.$$

The scheme of a MC simulation is now (see for example Ref. [29]) to approximate the integral (29) by its importance sampled average

$$\overline{A_q} = \frac{1}{M}\sum_{i=1}^{M} A_q(\Omega_i) \quad (34)$$

where the $M$ configurations $(\Omega_i, G_i)$ are distributed according to their statistical weight, i.e. with probability $P(\Omega_i,G_i) \propto W(\Omega_i,G_i)$.

The above considerations motivate the following MC algorithm:

1. Choose an arbitrary lattice, a site $x$ and a start configuration $([\Omega], G)$ of fields and bonds (the latter exhibiting a tree structure).

2. Generate (locally, i.e. by changing the value of a single, randomly chosen $\omega$) a new field configuration $([\Omega'], G)$. Calculate the weight ratio $\frac{W(\Omega',G)}{W(\Omega,G)}$ and accept the new configuration with probability $p_{site} = \min(1, \frac{W(\Omega',G)}{W(\Omega,G)})$.

3. Choose randomly a nearest-neighbor pair $(r_1, r_2)$ where $r_1$ and $r_2$ are *not* linked with each other by an occupied bond. Generate a new tree configuration $G'$ by breaking the bond between $r_2$ and its predecessor and linking $r_1$ and $r_2$ if $r_2$ is *not* an "ancestor" of $r_1$, which means that the path from $r_1$ towards $x$ along occupied bonds does not visit $r_2$. This procedure is illustrated in figure 1; it guarantees that the new bond configuration is again of tree type. Calculate the weight ratio of the new and the old bond configuration,

$$\frac{W(\Omega,G')}{W(\Omega,G)} = \frac{\omega(r_1)}{\omega(\text{predecessor of }r_2)}$$

and accept the new configuration with probability $p_{bond} = \min(1, \frac{W(\Omega,G')}{W(\Omega,G)})$. Note that all possible tree configurations can be attained within this procedure.

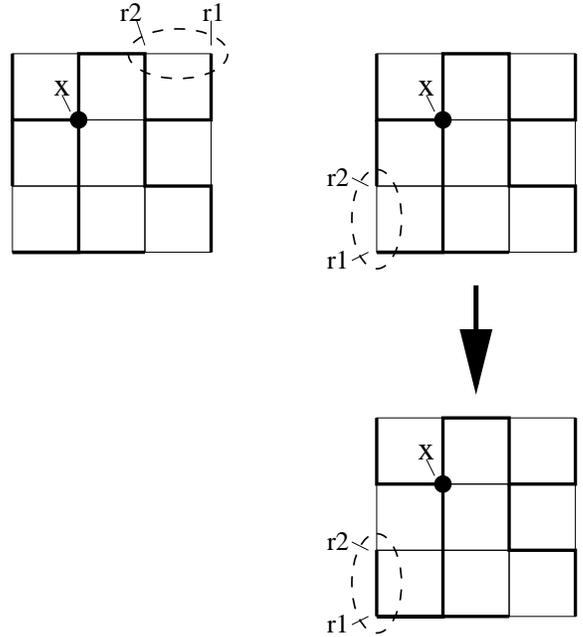

FIG. 1. *Example of contributing bond configurations which represent trees emerging from the site $x$. Left: $r_2$ is an ancestor of $r_1$; right: $r_2$ is no ancestor of $r_1$, therefore a new tree configuration evolves.*

Before I present the results of numerical simulations, it seems reasonable to get first a qualitative understanding of the model (29)-(33). I will discuss this point taking the one-dimensional chain as an example.

## B. Qualitative Discussion for a 1d chain

In a one-dimensional chain with $N$ lattice sites the sum over trees reduces to a triviality since only one tree, namely the chain itself, contributes. We will therefore omit the index $\{G\}$ in the sequel.

Consider first the limit $\gamma \to \infty$. In this case the factors $L_\gamma(\omega_i, \omega_j)$ in the weight $W(\Omega)$ force all $\omega_i$ to take roughly the same value which I denote by $\bar{\omega}$. Consequently one can approximate

$$\sqrt{\frac{\gamma}{2\pi}} L_\gamma(\omega_i, \omega_j) \simeq \bar{\omega} \delta(\omega_i - \omega_j) ,$$

and up to factors which are independent of $\bar{\omega}$ one finds

$$W(\Omega) \stackrel{\gamma \to \infty}{\Longrightarrow} const \times \bar{\omega}^{N-1} \prod_{i \smile j} \delta(\omega_i - \omega_j).$$

In the evaluation of the quantity $P_q$ all integrals except the integral over $\omega(x)$ break down due to the delta-function and one arrives at

$$P_q(\gamma) = \int_0^\infty d\bar{\omega} \exp(-N\bar{\omega}) \frac{\bar{\omega}^{q-2}}{(q-2)!} = N^{-(q-1)},$$

which is the correct limiting behavior.

For finite values of $\gamma$ the statistical mechanics of the model is governed by two competing effects: On the one hand, the variable $\omega_x$ takes values of order unity, since $\int D[\Omega] W(\Omega)$ contains the factor $\omega(x)^2 \exp -\omega(x)$. On the other hand, the radial coordinates at the boundaries of the chain, $\omega_1$ and $\omega_N$, say, show a strong singularity of type $1/\omega$ in eq. (32) and tend therefore to zero. The inverse square root of $\det A_x$ does not resolve this singularity. Depending on the value of $\gamma$ the intermediate sites "feel" the effect of the dynamics of $\omega_x$, $\omega_1$ and $\omega_N$ and balance it. For small $\gamma$ the effect of the $\omega$-"fixing" of the site $x$ is only weak, and consequently all $\omega_i$ – except $\omega_x$ – are pushed to zero. Thus, all IPR's $P_q$ tend to unity.

For lattices which are of non-tree type (for instance hypercubic lattices in $d \geq 2$) the qualitative picture is similar; in addition to the previous discussion any site $i$ with an (at the moment) large value of $\omega_i$ (compared to the other lattice sites) will accumulate occupied bonds at the cost of those sites $j$ with a small value of $\omega_j$. Conversely, a large number of occupied bonds at a given lattice site $i$ will favor a large value of $\omega_i$.

## C. The Calculation of the Determinant

This section deals with the "bottleneck" of the algorithm, namely the (technical, but numerically extremely important) question how to evaluate efficiently the determinant $\det A_x$, which appears in $W(\Omega, G)$, eq. (32). For the one-dimensional chain with $N$ sites it can be calculated with $\mathcal{O}(N)$ multiplications. However, for a hypercubic lattice of size $N = L \times L \times L$ and skew-symmetric (i.e. helical) boundary conditions [4], a naive implementation using standard libraries would take $\mathcal{O}(N^3) = \mathcal{O}(L^9)$ steps, which is clearly unacceptable. The reason is that although the matrix $A$ itself is sparse, it becomes in general dense after an $LU$-decomposition (which is used for the calculation of its determinant). Such a decomposition generates $\mathcal{O}(N^2)$ "fills" in the matrix, i.e. entries that were originally zero become non-zero after the $LU$-decomposition.

In a first improvement, one can modify the boundary conditions in such a way that the matrix $A$ displays a band structure with bandwidth $\mathcal{O}(L^2)$. This modification affects only a fraction of $\mathcal{O}(N^{-\frac{1}{3}})$ lattice sites – for which the boundary conditions are changed from periodic to free (hard) – and is therefore negligible in the thermodynamic limit. We can then use a standard Cholesky factorization for the determinant of a *real, symmetric, positive definite, banded* matrix. But since the CPU-time in this case is still of $\mathcal{O}(L^7)$ such algorithms are practicable only for very small systems ($L \lesssim 5$).

One can, however, calculate the determinant of $A_x$ in an approximate but controlled manner in $\mathcal{O}(L^3)$ time. First, one splits the matrix $A$ (28) into its diagonal and off-diagonal parts,

$$A = B - C, B_{ij} = a_i^{-1} \delta_{ij}, C_{ij} = \delta_{i \smile j} \equiv \begin{cases} 1 , & i \smile j \\ 0 , & otherwise \end{cases}.$$

For each matrix $T$, let $T_x$ denote the matrix which results from erasing the line and column $x$. Using the identity $\det = \exp \operatorname{tr} \ln$, and expanding the logarithm, one then has

$$\begin{aligned} \det A_x &= \det B_x \det(1 - B_x^{-1} C_x) \\ &= \det B_x \exp\left(-\sum_{n=1}^\infty \frac{1}{n} \operatorname{tr} M_x^n\right), \quad (35) \\ M_{ij} &= \sqrt{a_i a_j}\, \delta_{i \smile j} \end{aligned}$$

It remains to calculate $\operatorname{tr} M_x^n$ for $n = 1, 2, \ldots$ Note that $M_x$ does not display a simple structure, due to the effect of erasing one line and column. However, if we redefine

$$M_{ij} = \sqrt{a_i a_j}\, \delta_{i \smile j} \quad , \quad a_x := 0 \qquad (36)$$

(i.e. the line/column which is to be erased is replaced by a zero line/column) then it holds true that $\operatorname{tr} M_x^n = \operatorname{tr} M^n$.

---

[4] By skew-symmetric boundary conditions we mean that the lattice sites are numbered from 0 to $N-1$ where for each site $k$ its nearest neighbors are located at the positions $k \pm 1$, $k \pm L$ and $k \pm L^2$ (modulo $L^3$).

Thus, we need not bother about the somewhat unpleasant structure of $M_x$; instead, the problem of calculating $\det A_x$ is essentially reduced to the evaluation of

$$\operatorname{tr} M^n = \sum_i M^n_{ii} = \sum_{i \smile k_1 \smile k_2 \smile \ldots \smile k_{n-1} \smile i} a_i a_{k_1} a_{k_2} \ldots a_{k_{n-1}}$$

for positive $n$. The latter sum can be interpreted as a sum over all closed paths of length $n$ starting at (and returning to) a site $i$ on the underlying hypercubic lattice. Having this picture in mind, one clearly recognizes that

$$\operatorname{tr} M^{2n-1} = 0 \quad , \quad n = 1, 2, \ldots .$$

In addition, it can be shown that the series (35) converges at least like $\frac{1}{n} 2^{-n}$, which allows one to truncate the series after a finite value of $n$. In practice, $2n = 4$ is sufficient for an accuracy of better than 1%. Consequently, we can express the determinant of $A_x$ by

$$\det A_x \simeq \left( \prod_{y=1, y \neq x}^{N} a_i^{-1} \right) \exp - \left( \frac{1}{2} \operatorname{tr} M^2 + \frac{1}{4} \operatorname{tr} M^4 \right)$$

$$M_{ij} = \sqrt{a_i a_j} \, \delta_{i \smile j}, \ a_i = \frac{\omega_i}{2 \sum_{j \smile i} \omega_j}, \ a_x = 0. \quad (37)$$

Moreover, the remaining error largely cancels itself out when one calculates the ratio $\frac{W(\Omega', G)}{W(\Omega, G)}$, so that we end up with a relative accuracy of about $10^{-3}$. (In order to achieve a performance of $\mathcal{O}(L^3)$ essential multiplications in the calculation of $\det A_x$, it is, of course, necessary to use the sparsity of the matrix $M$ in the implementation of the matrix multiplications in eq. (37).

### D. Results of the Monte-Carlo Simulation

We have now gained a qualitative understanding of how the model works and how a numerical implementation is in principle done. I will therefore present the results of the above algorithm.

The algorithm has been checked at the example of a one-dimensional chain with $N$ sites, since exact results for $P_q(\gamma)$ are available for both $N = 2$ and $N = \infty$. The agreement between analytical and numerical results is very good.

From the analytical calculations of the model (10) on the BL [19] we expect the critical disorder $\gamma_c$ for the $3d$ hypercubic lattice in the range between $10^{-2}$ and $10^{-1}$. We focus our attention therefore on this order of magnitude of $\gamma$ and calculate $P_q(\gamma; L)$ for system sizes $L = 4, 5, 6, 7, 8, 10$. [5] This is done as follows. To obtain a set of $P_q(\gamma; L)$ (for fixed $\gamma$ and $L$, but variable $q = 2, 4, 6, \ldots$) we perform typically some $10^6 - 10^7$ Monte-Carlo steps (MCS), where each step represents a complete sweep i.e. consists of $N = L^3$ random updates of sites and bonds as described in section IV A. We store the pair $(\omega(x), \exp - \sum_{y \neq x} \omega(y))$ after each tenth MCS. From this data one can afterwards average (34) the expression $A_q(\Omega)$ (31) for each $q$ separately and obtains thus $P_q(\gamma; L)$ for all $q$. In order to make sure that the above data-points $A_q(\Omega_i)$ (where $i$ labels the MCS), which are to be averaged, are statistically independent, we calculate the autocorrelation function

$$C(n) = \overline{A_q(\Omega_i) A_q(\Omega_{i+n})} - \overline{A_q}^2 \quad (38)$$

and take for averaging only each $(N_{corr})^{th}$ data-point $A_q(\Omega_i)$. Here, $N_{corr}$ is defined as the length on which $C(n)$ has collapsed to a relative value of 10%. Given these $N_{eff} = N_{total}/N_{corr}$ data-points, the statistical error scales like $N_{eff}^{-\frac{1}{2}}$.

After we have produced for each $q$ and $L$ the "raw data" $y_i = P_q(\gamma_i; L)$ (for discrete $\gamma_i$) together with their statistical errors $\sigma_i$, this data is subjected to a $\chi^2$-fit procedure. This procedure yields a continuous function $P_q(\gamma; L)$ that interpolates between the values $y_i$ by a least squares fit (see for instance [30] or an appropriate textbook). As an illustration figure 2 shows the resulting set of functions for $q = 2$ and $L = 4, 5, 6, 7, 8, 10$.

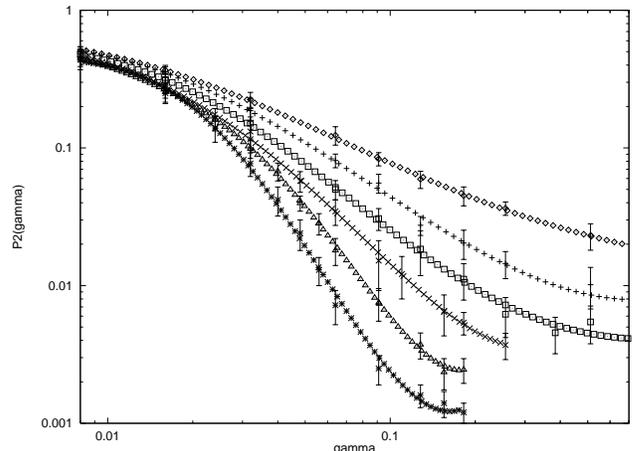

FIG. 2. The "raw data" $P_2(\gamma_i; L)$ together with their statistical errors, and the interpolating function $P_2(\gamma; L)$ as calculated from a $\chi^2$-fit. System sizes are $L = 4, 5, 6, 7, 8, 10$ (from top to bottom).

These functions tend in fact to the correct limits for $\gamma \to 0$ and $\gamma \to \infty$ and do not intersect each other.

### E. Multifractal Analysis

The goal of this section is to investigate the expected multifractal structure of the MIT, by extracting from the set of functions $P_q(\gamma; L)$ the generalized dimensions $d^*(q)$ (18) and the correlation length exponent $\nu$ via a finite size analysis.

---

[5] It turns out to be somewhat disadvantageous that the precise value of $\gamma_c$ is not yet known which hampers a direct investigation of the critical regime right from the beginning.

We determine $\gamma_c$ and the set $\tau(q)$ (from which $d^*(q)$ follows immediately) by multiplying each function $P_q(\gamma; L)$ with a factor $L^{\tilde{\tau}(q)}$ – with a "guess" $\tilde{\tau}(q)$. If (and only if) $\tilde{\tau}(q) = \tau(q)$ then all curves $L^{\tilde{\tau}(q)} P_q(\gamma; L)$ (for fixed $q$ and variable $L$) intersect in one single point at a nontrivial value $\gamma = \gamma_c$. This is due to the expected 1PS behavior in the critical regime (where $\xi \gg l_{mic}$),

$$P_q(\gamma; L) \propto L^{-\tau(q)} f\left((\gamma - \gamma_c) L^{\frac{1}{\nu}}\right) , \qquad (39)$$

compare eq. (18). ($\tau(q)$ does not depend on $\gamma$.) Usually, the above "fitting" procedure is not very accurate, since the critical coupling constant $\gamma_c$ is not known and therefore *two* parameters ($\tau(q)$ and $\gamma_c$) are determined simultaneously, which might yield quite a large uncertainty. Note, however, that $\gamma_c$ must take the same value for all $q$, therefore this approach allows for a rather precise estimation of $\gamma_c$. Thus, for each $q$, effectively only one parameter, $\tau(q)$, is fitted. (Once the value of $\gamma_c$ is known, one could alternatively extract the set $\tau(q)$ from a plot of $\ln P_q(\gamma_c; L)$ versus $\ln L$,

$$\ln P_q(\gamma_c; L) = -\tau(q) \ln L + const$$

which is a straight line with (negative) slope $\tau(q)$.)

Proceeding along this scheme we observe indeed 1PS and multifractality at the critical point: the dimensions $d^*(q) = \tau(q)/(q-1)$ are not constant but fall off with increasing $q$, approaching a saturation value $d_\infty \simeq 0.3$. In figure 3 we show as an example the 1PS behavior for $q = 4$. Here, we have multiplied the function $P_4(\gamma, L)$ with a factor $\tau(4) \simeq 1.8$ which results (roughly) in a single intersection point around $\gamma_c \simeq 0.04$.

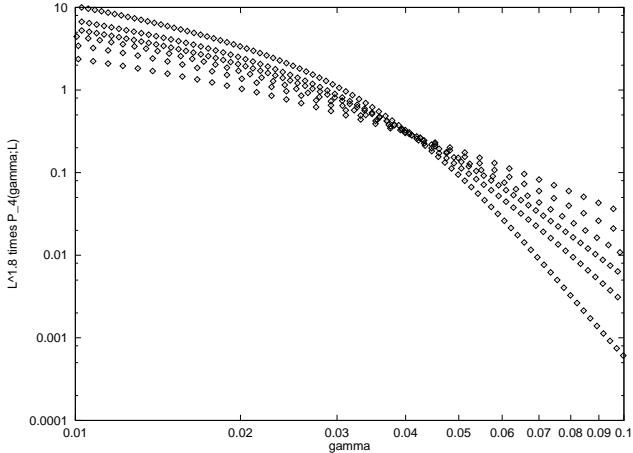

FIG. 3. The curves $L^{\tau(q)} P_q(\gamma; L)$ for $q = 4$, $L = 4, 5, 6, 7, 8, 10$ and $\tau(4) = 1.8$.

We find

$$\gamma_c = 0.038 \pm 0.004 .$$

where for different $q$ the numerically obtained values for $\gamma_c$ fluctuate a little bit around that value. The set $\tau(q)$ is displayed in figure 4.

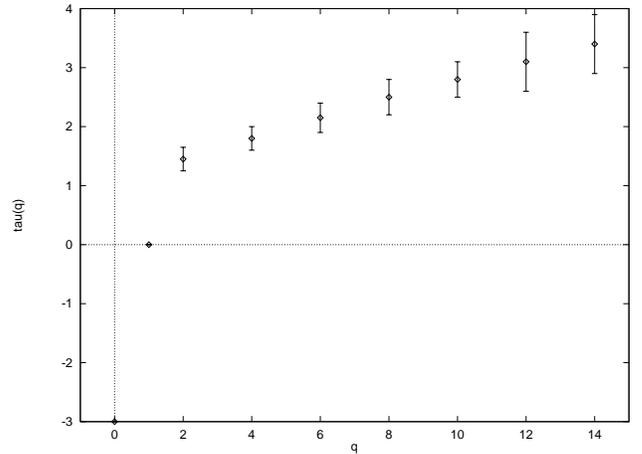

FIG. 4. The obtained values for $\tau(q)$ with $q = 2, 4, 6, 8, 10, 12, 14$. The trivial results $\tau(0) = -d$ and $\tau(1) = 0$ are added.

In principle we can from this data determine the so-called $f(\alpha)$ spectrum, [25]

$$f(\alpha) = \alpha(q) q - \tau(q) , \quad \alpha(q) = \frac{d\tau(q)}{dq} . \qquad (40)$$

In our approach, however, $q$ is not a continuous variable, as we do not calculate the quantum mechanical wave function (from which $P_q$ for *any* value of $q$ could be deduced). Instead, $q$ takes only discrete (and even) values. We therefore omit the calculation of the $f(\alpha)$ spectrum, since the regime $\alpha \simeq \alpha_0$ (the maximum of $f(\alpha)$) cannot be probed by our approach.

The localization length exponent $\nu$ can be obtained if one makes a guess $\tilde{\nu}$ and plots $P_q(\gamma; L) L^{\tau(q)}$ versus $L^{1/\tilde{\nu}}$ for $\gamma \simeq \gamma_c$ and fixed $q$. For, from eq. (39) one recognizes that, if $\tilde{\nu} = \nu$, then the curves $P_q(\gamma; L) L^{\tau(q)}$ have for different $L$ (but fixed $q$) the same dependence on $\gamma - \gamma_c$ and fall together, as long as the criticality condition $\xi \gg l_{mic}$ is satisfied.

We obtain

$$\nu = 1.15 \pm 0.15 ,$$

compare figure 5.

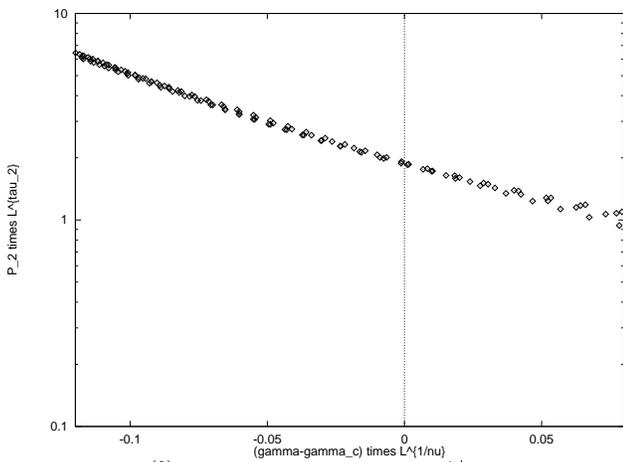

FIG. 5. $L^{\tau(2)} P_2(\gamma; L)$ vs. $(\gamma - \gamma_c) L^{1/\nu}$ ($\tau(2) = 1.45$, $\nu = 1.15$) for different $L$.

Although this value of $\nu$ (and the findings for $d^*(q)$) is surely not the result of a high-precision calculation, it fits well into the "landscape" of estimates for $\nu$ for the 3$d$ Anderson model that have been obtained in the past by means of several different methods, see for instance [31–35]. These values range – roughly – from $\nu \simeq 0.97$ to $\nu \simeq 1.5$ together with some quite strongly deviating estimates. It is, of course, not guaranteed that the localization exponents of our toy model and of the Anderson model coincide exactly. But the qualitative agreement of its numerical values is a further indication that the above toy model yields a qualitatively correct description of the localization transition.

### F. Limitations of the Algorithm

Let us briefly discuss the limitations of the applicability of the above outlined algorithm. A natural limitation is given by a maximal system size $L \lesssim 15$, because of restrictions in CPU time. Further, because of the "critical slowing down" one needs more and more MCS to investigate the critical point. Apart from that, there are other restrictions which are specific to our model.

First, for $\gamma \to 0$, all $\omega_i$ tend to zero, but *non-uniformly*. Thus, the exponent of $L_\gamma$ (33) fluctuates strongly – in spite of the smallness of $\gamma$ – due to the "almost-singularity" which appears in $\omega_i/\omega_j$. Consequently, the weight ratio $W(\Omega', G)/W(\Omega, G)$ might take very small or large values and produce a numerical underflow or overflow. This problem can be resolved if one chooses in each MCS only small deviations from the previous configuration so that the large numbers cancel each other mostly. This is, of course, at the cost of CPU time, since one needs more MCS to "scan" the configuration space. Fortunately, these "precursors" of the singularity at $\gamma = 0$ do not affect the present analysis too seriously, since $\gamma_c$ is still large enough so that simulations around $\gamma \simeq \gamma_c$ are not plagued by this feature.

More problematic seems the structure of the model at the opposite limit, that is, for $\gamma \gtrsim \gamma_c$ and large $L$. As the symmetry breaking term $A_q$ (31), which is to be averaged, is essentially of the form $\exp - \sum_i \omega_i$, the fluctuations of this quantity become enormously large if $L$ is large and the $\omega_i$ are not small, which means that $\gamma$ is not small. Therefore one needs a large number of MCS, since – roughly speaking – the distribution of $A_q$ is essentially of log-normal type, and thus the less probable values dominate its average.

One might ask whether this problem cannot be circumvented if – speaking in general words – instead of averaging a strongly fluctuating quantity $A_i$ according to a weight distribution $W_i$ one averages, say, $\sqrt{A_i}$ according to $W_i \sqrt{A_i}$,

$$\overline{A} \equiv \frac{\sum_i A_i W_i}{\sum_i W_i} \mapsto \frac{\sum_i A_i^{\frac{1}{2}} V_i}{\sum_i A_i^{-\frac{1}{2}} V_i} \, , \; V_i \propto W_i A_i^{\frac{1}{2}}$$

and hopes that the fluctuations are damped by the square root. However, numerical tests have shown, that this procedure yields no improvement. Instead, the problem of exponentially large fluctuations seems to be inherent to the MIT (at least in the formulation of a non-linear $\sigma$-model).

It is the last restriction that hampers me from investigating systems with $L \gtrsim 13$, say. However, as we have seen, the 1PS sets in already for $L \lesssim 10$ so that the investigation of still larger systems seems to be unnecessary.

### V. DISCUSSION

In this work, we have put forward a non-linear $\sigma$-toy-model which we assume to give a qualitatively correct description of the localization transition. After a short discussion of a simple case, in which this model is exactly solvable, we outlined an MC-algorithm and provided numerical data which demonstrate that the localization transition within the supersymmetric formulation is a continuous phase transition which can be described by 1PS. The multifractal nature of the transition has been confirmed.

It would be an interesting future task to transfer the MC algorithm from the toy model of the present work onto the supermatrix models, for instance onto the unitary model. It seems that the additional degrees of freedom, that arise in such a transfer, are not too difficult to handle.[6] The essential features of these models are already encountered in the present MC algorithm. Such a transfer would be advantageous in the sense that a direct connection between physical quantities and correlation functions of the model is available. Moreover, one can calculate the IPR's $P_q$ for *all* (positive integer) $q$ not only for $q$ even.

---

[6] However, due to the presence of more than one Grassmann variables together with their conjugates, the contributing bond configurations in section IV are no more necessarily of tree type.

One could also utilize the above MC algorithm in a very similar way in order to calculate the order parameter function $F(\psi^r)$, eq. (26) in $d = 3$. This yields directly the distribution function of the local amplitudes of wave functions [36,37]) instead of each of its moments. Further, one can investigate the case $d = 2$ (see for example Ref. [38]) where it is possible to simulate systems of larger (linear) size than in three dimensions.

I am very grateful to M.R. Zirnbauer for numerous stimulating and encouraging discussions from which I have benefited very much. Further, I acknowledge with thanks helpful discussions with A. Altland, B. Huckestein, M. Janssen, and H. Rieger. Finally, I wish to thank the German National Scholarship Foundation for financial support.